# On Constancy of the Speed of Light in the Global Positioning System

Yang-Ho Choi

Department of Electrical and Electronic Engineering
Kangwon National University
Chunchon, Kangwon-do, 200-701, South Korea

**Abstract:** The speed of light in the global positioning system (GPS) is theoretically investigated through the analysis of navigation equations together with metric tensors for light speed constancy. The GPS equations employ a Galilean transformation between the Earth centered inertial (ECI) and the Earth centered Earth fixed (ECEF) frames. This paper clearly shows that under the nonrelativistic transformation, the speed of light inevitably becomes anisotropic in the ECEF frame if it is isotropic in the ECI frame. One can make the speed of light appear isotropic by introducing the standard synchronization into the ECEF. It is well known that the Sagnac correction due to the Earth's rotation should be made for accurate positioning. However the Sagnac correction comes to be cancelled out by the standard synchronization so that GPS positioning would become less accurate. Using an exact coordinate transformation between inertial and rotating frames presented recently, we make further investigation on speeds of light in a rotating plate as well as in the GPS. The Galilean transformation turns out to be a first-order approximation to the exact one, which explains why GPS positioning has worked so well based on it. In the experiment of the Sagnac effect on a rotating plate, the speed of light is also shown to be anisotropic. The introduction of the standard synchronization for the light speed constancy can be considered. However, it has been shown that two counter-rotating light beams travel the same distance in the standard-synchronized rotating frame so that the Sagnac effect cannot occur if their speeds are really equal. The occurrence of the Sagnac effect means the anisotropy of the light speed in the rotating frame. Moreover the introduction of the synchronization causes the problem of time gap that multiple times are defined at the same place, as explicitly presented.
  .
**Keywords:** Light speed constancy, Rotating frame, Standard synchronization, GPS, Sagnac correction.

# I. INTRODUCTION

The global positioning system (GPS) has been widely utilized in a variety of applications such as navigation, synchronization in mobile communications, position tracking, and geographic information system (GIS). Synchronization of clocks that are located at different places on the Earth can be globally made using the GPS. Moreover GPS users can readily discover their positions at any place if at least four GPS signals can be received. It is well known that the GPS has worked very accurately. According to the official US government information about the GPS [1], "Real-world data from the FAA show that their high-quality GPS SPS receivers provide better than 3.5 meter horizontal accuracy." In order for GPS users to extract their precise times and positions from high accurate information that GPS can provide, relativistic effects including gravitational ones should be accounted for [2–6].

Navigation equations exploit the positions and times of the GPS satellites in the Earth centered inertial (ECI) frame, which is assumed to be isotropic. Those of GPS users are represented in the Earth centered Earth fixed (ECEF) frame. Hence relativistic relationship between the ECI and ECEF frames comes into play. The GPS positioning implicitly needs the one-way speed of light in the ECEF frame, though it has been often considered as a non-measurable quantity because of the problem of synchronization between two spatially separated clocks [7]. Hence GPS measurements can provide empirical information on whether or not the one-way speed of light is really isotropic in the ECEF frame. This paper theoretically investigates the one-way speed of light based on navigation equations and metric tensors for light speed constancy.

The speed of light is clearly shown to be anisotropic in the ECEF frame. Navigation equations utilize a Galilean transformation in cylindrical coordinates between the ECI and ECEF frames. As a result, the light speed inevitably becomes anisotropic in the ECEF frame if it is isotropic in the ECI. The metric tensor can be introduced into the ECEF frame for the standard synchronization [7, 8] between spatially separated clocks so that the speed of light appears to be $c$ where $c$ is the speed of light in vacuum. It is well known that the so called Sagnac correction should be made to extract accurate positioning information from GPS signals received [2–6]. However, as soon as the standard synchronization is applied, the term of the Sagnac correction is cancelled out, thereby resulting in less accurate positioning, which implies the anisotropy of the speeds of GPS electromagnetic signals in the ECEF.

Using the exact space-time coordinate transformation between inertial and rotating frames presented recently [9], we further examine speeds of light in a rotating plate as well as in the GPS. The Galilean transformation employed by GPS equations turns out to be a first-order approximation to the exact one, which validates the use of the nonrelativistic transformation. In the experiment of the Sagnac effect on a rotating plate, the speed of light is also anisotropic in the rotating frame. If one makes an attempt to introduce the standard synchronization into the rotating frame for light speed constancy, it would be, however, impossible on account of the problem of time gap that multiple times are defined at the same place, as explicitly presented in the paper. Even if it is carried out, the two counter-rotating light beams travel the same distance in the standard-synchronized rotating frame so that fringe shifts cannot occur.

Though the information on the times and positions in the ECI frame that the GPS satellites send may contain some errors due to various effects [4] which deviate from a presumed model, it is assumed that they are negligible. Good working of the GPS implies that the information is sufficiently accurate. It would be meaningless to examine the speed of light with the GPS information if it is erroneous. Furthermore, the objective of this paper is to make investigation of light speed constancy in the ECEF frame, a rotating frame, in an ideal situation and thus the other effects that affect the speed of light except the Earth's rotation are not considered.

## II. GPS NAVIGATION EQUATIONS

In this section the coordinate systems $S$ and $\tilde{S}$ denote the ECI and ECEF frames, respectively, and their spatial coordinate vectors are represented as $\boldsymbol{p} = [r, \phi, z]^T$ and $\tilde{\boldsymbol{p}} = [\tilde{r}, \tilde{\phi}, \tilde{z}]^T$ where $T$ stands for the transpose. In case a Galilean transformation in cylindrical coordinates between $S$ and $\tilde{S}$ is employed, their coordinates are related by

$$\tilde{t} = t, \quad \tilde{r} = r, \quad \tilde{\phi} = \phi - \omega \tilde{t}, \quad \tilde{z} = z \tag{2.1}$$

where $\omega$ is the angular speed of the Earth relative to $S$. The space-time interval in $\tilde{S}$ can be written as

$$ds^2 = -c^2 dt^2 + dr^2 + r^2 d\phi^2 + dz^2. \tag{2.2}$$

The space-time of $S$ is assumed to be isotropic so that $ds = 0$ for electromagnetic signals. Substituting (2.1) into (2.2) yields

$$ds^2 = -\gamma_{\tilde{\beta}}^{-2} c^2 d\tilde{t}^2 + 2c\tilde{\beta}\tilde{r} d\tilde{\phi} d\tilde{t} + d\tilde{l}^2 \tag{2.3}$$

where

$$\tilde{\beta} = c^{-1}\tilde{r}\omega \tag{2.4}$$

$$\gamma_\alpha = (1 - \alpha^2)^{-1/2} \tag{2.5}$$

$$d\tilde{l}^2 = d\tilde{r}^2 + \tilde{r}^2 d\tilde{\phi}^2 + d\tilde{z}^2. \tag{2.6}$$

It is worth noting that $d\tilde{l}$ is equivalent to the distance when the space of $\tilde{S}$ is Euclidean.

A GPS user, who is at rest in the ECEF, receives electromagnetic signals from $N$ GPS satellites.

The $n$th one is located at $\boldsymbol{p}_n = [r_n, \phi_n, z_n]^T$ in the ECI when it transmits an information signal at a time instant $t = t_n$. GPS positioning begins with the navigation equations [4–7]

$$c^2(t_0 - t_n)^2 = \|\boldsymbol{p}_0 - \boldsymbol{p}_n\|^2, \quad n = 1, \cdots, N \tag{2.7}$$

where $N \geq 4$, $\|\cdot\|$ designates the Euclidean norm, and $\boldsymbol{p}_0$ is the spatial coordinates in $S$ of the user at the instant $t = t_0$ that the transmitted signals are received. The electromagnetic waves carry information on the positions and times of GPS satellites. As $ds^2 = 0$ for electromagnetic signals, from (2.3), we have

$$-\gamma_{\tilde{\beta}}^{-2} c^2 d\tilde{t}^2 + 2c\tilde{\beta}\tilde{r}d\tilde{\phi}d\tilde{t} + d\tilde{l}^2 = 0. \tag{2.8}$$

Equation (2.8) is a quadratic equation of $d\tilde{t}$. Within the first-order approximation of $\tilde{\beta}$, the solution to (2.8) with respect to $d\tilde{t}$ can be expressed as [5, 6]

$$c d\tilde{t} = d\tilde{l} + d\tilde{\xi} \tag{2.9}$$

where

$$d\tilde{\xi} = \tilde{\beta}\tilde{r}d\tilde{\phi}. \tag{2.10}$$

The term $d\tilde{\xi}$ is called the Sagnac correction. The spatial vectors $\boldsymbol{p}_n$, $n = 1, \cdots, N$, in $S$ are denoted as $\tilde{\boldsymbol{p}}_n$ in $\tilde{S}$. Suppose that the GPS user, whose position is $\tilde{\boldsymbol{p}}_0$ in $\tilde{S}$ at the instant $\tilde{t} = \tilde{t}_0$, receives the $n$th GPS signal after $\Delta\tilde{t}_n \,(= \tilde{t}_0 - \tilde{t}_n)$ has passed since the transmission of $A_n$ at $\tilde{\boldsymbol{p}}_n$. Then the travel time $\Delta\tilde{t}_n$ of the transmitted signal from $A_n$ to the user is written from (2.9) as

$$\Delta\tilde{t}_n = c^{-1}\int_{\tilde{\boldsymbol{p}}_n}^{\tilde{\boldsymbol{p}}_0} d\tilde{l} + c^{-1}\int_{\tilde{\boldsymbol{p}}_n}^{\tilde{\boldsymbol{p}}_0} d\tilde{\xi}, \quad n = 1, \cdots, N. \tag{2.11}$$

Since $\tilde{t} = t$, $\Delta\tilde{t}_n$ is identical with the elapsed time $\Delta t_n$ during the propagation of the signal in $S$.

### III. LIGHT SPEED CONSTANCY AND STANDARD SYNCHRONIZATION
### 1. Navigation Equations Per se

The coordinate transformation (2.1) between $S$ and $\tilde{S}$ is nonrelativistic, and there are not essentially any relativistic things associated with special relativity or general relativity as for as the navigation equations (2.7) or (2.11) are concerned. As the invariant interval that is a basic asset of the theories of relativity is utilized as in (2.8) together with the well-known Langevin metric usually used for the relativistic analysis of circular motion, the resulting results may seem relativistic so that the speed of light may be considered to be $c$ in $\tilde{S}$ regardless of its direction. However, (2.3) is nothing more than the representation of $ds^2$ in $\tilde{S}$. Equation (2.8), which is not associated in itself with the standard synchronization for light speed constancy, results from the fact that $ds^2 = 0$ in $S$ for electromagnetic signals.

Basically $\Delta \tilde{t}_n \ (= \Delta t_n)$ and $\tilde{p}_0$ can be solved by two slightly different, but actually same, computational methods. First method is to directly solve (2.7) so that $\Delta t_n$ and $p_0$ are found in $S$. Then $p_0$ is converted into $\tilde{S}$ according to (2.1). Alternatively, the spatial vectors $p_n$ are first converted into $\tilde{S}$ to obtain $\tilde{p}_n$, and then $\Delta \tilde{t}_n$ and $\tilde{p}_0$ are found by solving (2.11). It is obvious that the speed of light is anisotropic in $\tilde{S}$ with respect to coordinate time $\tilde{t}$ if it is isotropic in $S$.

## 2. Introduction of Metric Tensors

One can make the speed of light appear to be $c$ by introducing metric tensors. The square of a differential line element is generally written as

$$ds^2 = g_{\alpha\beta} dx^\alpha dx^\beta \\
= g_{00}(dx^0)^2 + 2g_{0i} dx^i dx^0 + g_{ij} dx^i dx^j \tag{3.1}$$

where the tensor element $g_{00}$ is less than zero and repeated Greek and Latin indices are summed over 0 through 3, and over 1 through 3, respectively. Equation (3.1) can be rewritten as

$$ds^2 = g_{00}(dx_\bullet^0)^2 + dl_\circ^2 \tag{3.2}$$

where

$$dx_\bullet^0 = dx^0 + g_{00}^{-1} g_{0i} dx^i \tag{3.3}$$

$$dl_\circ = (dl^2 - g_{00}^{-1} g_{0i} g_{0j} dx^i dx^j)^{1/2} \tag{3.4}$$

with $dl = (g_{ij} dx^i dx^j)^{1/2}$ representing the spatial distance in the coordinate system.

There appear different kinds of time in (3.3). The time $x^0$ and $x_\bullet^0$ denote the coordinate time (CT) and the resynchronized time, respectively, normalized with respect to $c$. Here the latter will be called adjusted time (AT). There is another important time, the proper time (PT). The normalized differential PT is defined as [8]

$$dx_\circ^0 = |ds|. \tag{3.5}$$

In case the object of interest is at rest in the frame so that $dx^i = 0$, $i = 1, \cdots, 3$, (3.5) is expressed as

$$dx_\circ^0 = |g_{00}|^{1/2} dx^0. \tag{3.6}$$

For simplicity, we omit the word "normalized" so that for example, CT can mean either $x^0/c$ or $x^0$.

The metric tensor for the invariance of space-time intervals can be employed in $\tilde{S}$ so that $ds^2 = d\tilde{s}^2$ where $d\tilde{s}$ is the space-time interval in $\tilde{S}$. When $(x^0, x^1, x^2, x^3) = (c\tilde{t}, \tilde{r}, \tilde{\phi}, \tilde{z})$ the squared line element $d\tilde{s}^2$ is expressed as (2.3) under the invariance, which leads to

$$g_{00} = -\gamma_{\tilde{\beta}}^{-2}, \quad g_{02} = g_{20} = \tilde{\beta}\tilde{r}, \quad g_{11} = g_{33} = 1, \quad g_{22} = \tilde{r}^2 \tag{3.7}$$

and the other elements are all zero. Substituting (3.7) into (3.3) and (3.4) yields

$$cdt_\bullet = cdt - \gamma_{\tilde{\beta}}^2 \tilde{\beta}\tilde{r}d\tilde{\phi} \tag{3.8}$$

$$dl_\circ = (dl^2 + \gamma_{\tilde{\beta}}^2 (\tilde{\beta}\tilde{r}d\tilde{\phi})^2)^{1/2} \tag{3.9}$$

where $dl = d\tilde{l}$. Within the first order approximation of $\tilde{\beta}$, $\gamma_{\tilde{\beta}}$ is given by

$$\gamma_{\tilde{\beta}} = 1 \tag{3.10}$$

and (3.8) and (3.9) are approximated as

$$cdt_\bullet = cdt - \tilde{\beta}\tilde{r}d\tilde{\phi} \tag{3.11}$$

$$dl_\circ = dl. \tag{3.12}$$

Recall $\tilde{t} = t$. Putting (2.9) and (3.12) in (3.11), we have

$$cdt_\bullet = dl_\circ. \tag{3.13}$$

If the distance is defined as (3.4) and $\gamma_{\tilde{\beta}} = 1$, the speed of light in $\tilde{S}$ becomes $c$ with respect to AT, not CT. The Sagnac correction is needed because the speed of light with respect to CT is anisotropic

## 3. Various Kinds of Time and Distance and Speeds of Light

It is necessary to clearly see the meaning of (3.3). Consider two points $p_0$ and $p_1$ where $p_m = (x_m^1, x_m^2, x_m^3)$, $m = 0, 1$, $p_1 = p_0 + dp$, and $dp = (dx^1, dx^2, dx^3)$. A symbol $W_m$ designates the world line corresponding to $p_m$. According to (3.3), we can set AT over $W_0$ equal to CT, i.e.,

$$x_\bullet^0 = x^0 \tag{3.14}$$

and then over $W_1$

$$x_\bullet^0 = x^0 + g_{00}^{-1} g_{0i} dx^i. \tag{3.15}$$

At first sight (3.14) and (3.15) seem to contract each other. It needs to understand their meaning. Equations (3.14) and (3.15) represent the relationships between CT and AT over $W_0$ and $W_1$, respectively. The seeming contradiction arises when $x_\bullet^0$ and $x^0$ in (3.14) are considered to equal respective ones in (3.15) at the same time. Both of them in (3.14) cannot be equal to respective ones in (3.15) at the same time and one of them is different from the corresponding one. For example, if $x_\bullet^0$ in (3.14) equals $x_\bullet^0$ in (3.15), $x^0$ in (3.14) becomes different from $x^0$ in (3.15). Then $x^{0(0)} = x_\bullet^0$ and $x^{0(1)} = x_\bullet^0 - g_{00}^{-1} g_{0i} dx^i$ where $x^{0(m)}$ denotes CT over $W_m$. Fig. 1 illustrates $x_\bullet^{0(m)}$ with respect to $x^0$ where $x_\bullet^{0(m)}$, $m = 0, 1$, denotes AT over $W_m$. The black dot on $W_1$ indicates $x_\bullet^{0(1)}$ which has the same value as $x_\bullet^{0(0)}$. Noting $x_\bullet^{0(1)} = x_\bullet^{0(0)} = x^{0(0)}$ and using (3.15), the CT time $x^{0(1)}$ corresponding to the $x_\bullet^{0(1)}$ is given by $x^{0(1)} = x^{0(0)} - g_{00}^{-1} g_{0i} dx^i$. In fact, the black dot on $W_1$ indicates $x^0 (= x^{0(1)})$ corresponding to $x_\bullet^{0(1)}$.

As illustrated in Fig. 2, an object leaves $p_0$ at $x^0 = x_0^{0(0)}$ and, at $x^0 = x_1^{0(1)}$, reaches $p_1$, from which it returns back to $p_0$ at $x^0 = x_2^{0(0)}$. Let $x_{\bullet k}^{0(m)}$ be the AT corresponding to $x_k^{0(m)}$ where $k = 0, \cdots, 2$ and $m = 0, 1$. Time difference in CT is denoted as

$$\Delta x_{kl}^{0(m)} = x_k^{0(m)} - x_l^{0(m)} \tag{3.16}$$

and in AT

$$\Delta x_{\bullet kl}^{0(m)} = x_{\bullet k}^{0(m)} - x_{\bullet l}^{0(m)} \tag{3.17}$$

where $k, l = 0, \cdots, 2$. Hereafter, suppose that $x_{\bullet k}^{0(0)} = x_{\bullet k}^{0(1)}$ in this subsection. And in case a symbol for CT or AT time (or time interval) has no superscript specifying the world line, say $x^0$, its world line is $W_0$, say $x^{0(0)}$, or it is the same irrespective of world lines, say $x^{0(0)} = x^{0(1)}$. It is easy to see from (3.14) and (3.15) that given $x_{\bullet k}^0 (= x_{\bullet k}^{0(0)} = x_{\bullet k}^{0(1)})$, the CT differences are the same, i.e.,

$$\Delta x_{kl}^{0(m)} = \Delta x_{\bullet kl}^0. \tag{3.18}$$

One can also see the relationship (3.18) from Fig. 2, recalling the black dots indicate CT times corresponding to their respective AT times. As the CT interval is related to the AT interval by (3.18), the PT interval can be expressed from (3.6) as

$$dx_\circ^0 = | g_{00} |^{1/2} dx_\bullet^0. \tag{3.19}$$

In case the travelling object is a light particle, $ds^2 = 0$. Setting (3.2) equal to zero and solving the resulting quadratic equation of $dx^0$, we have

$$dx_+^0 = \frac{g_{0i} dx^i + | g_{00} |^{1/2} dl_\circ}{| g_{00} |} \tag{3.20}$$

$$dx^0_- = \frac{g_{0i}dx^i - |g_{00}|^{1/2} dl_\circ}{|g_{00}|}. \tag{3.21}$$

Then $x^0_0 = x^0_1 + dx^0_-$ and $x^0_2 = x^0_1 + dx^0_+$ [8]. The departure AT time at $p_0$ is given from (3.14) by $x^0_{\bullet 0} = x^0_0$. When the standard synchronization is employed for synchronizing clocks at different places, the arrival time $x^0_{s1}$ of the light particle at $p_1$ is written as

$$x^0_{s1} = \frac{x^0_0 + x^0_2}{2} = x^0_1 - \frac{g_{0i}dx^i}{|g_{00}|}. \tag{3.22}$$

The $x^0_{s1}$ is the same as $x^{0(1)}_{\bullet 1}$ given in (3.15), which means that AT is the time according to the standard synchronization. The round trip time of the light particle is $\Delta x^0_{20}$, which is given by

$$\Delta x^0_{20} = x^0_2 - x^0_0 = 2|g_{00}|^{-1/2} dl_\circ. \tag{3.23}$$

From (3.18) $\Delta x^0_{\bullet 20} = \Delta x^0_{20}$. If the speed of light is $c$ under the standard synchronization, the distance $dl_\bullet$ between $p_0$ and $p_1$ should be equal to $\Delta x^0_{20}/2$ and thus

$$\begin{aligned} dl_\bullet &= |g_{00}|^{-1/2} dl_\circ \\ &= |g_{00}|^{-1/2} (dl^2 - g_{00}^{-1} g_{0i} g_{0j} dx^i dx^j)^{1/2} \end{aligned} \tag{3.24}$$

In consequence, if clock synchronization is made as (3.3) and spatial distance is defined as (3.24), the speed of light appears to be $c$ with respect to AT.

The line element $ds^2_\bullet$, which is defined as $ds^2_\bullet = -(dx^0_\bullet)^2 + dl^2_\bullet$, is written from (3.19), (3.24) and (3.2) as

$$ds^2_\bullet = |g_{00}^{-1}| ds^2. \tag{3.25}$$

Clearly if $ds^2 = 0$, $ds^2_\bullet = 0$, which is consistent with the approximate relationship (3.13) where $dl_\circ = dl_\bullet$ as $\gamma_{\tilde\beta} = 1$. Without any approximations, light speed constancy holds with respect to AT in the standard-synchronized coordinate system. However the invariance of the interval does not for $|g_{00}| \neq 1$, as shown in (3.25). PT and $dl_\circ$ correspond to scaled AT and $dl_\bullet$, as can be seen from (3.19) and (3.24). It is easy to see that

$$ds^2 = -dx_\circ^2 + dl_\circ^2. \tag{3.26}$$

In case PT and $dl_\circ$ are employed for clock time and spatial distance, the invariance of the interval as well as the light speed constancy is maintained.

The round trip speed of a light particle that travels between $p_0$ and $p_1$ is $c$ also with respect to CT, which results from the definition of distance that $dl_\bullet = \Delta x^0_{20}/2$. However, the one-way speeds are other than $c$. Let us investigate one-way speeds in terms of CT. The photon which leaves $p_0$ when $x_\bullet^0 = x_{\bullet 0}^0$ arrives at $p_1$ when $x_\bullet^0 = x_{\bullet 1}^0$, and is reflected back to $p_0$ when $x_\bullet^0 = x_{\bullet 2}^0$, as shown in Fig. 2. Recalling that $\Delta x_{\bullet 10}^0 = \Delta x_{\bullet 21}^0 = dl_\bullet$ and using (3.14) and (3.15), we have

$$\Delta x_{10}^{0(10)} = dl_\bullet - g_{00}^{-1} g_{0i} dx^i \tag{3.27}$$

$$\Delta x_{21}^{0(01)} = dl_\bullet + g_{00}^{-1} g_{0i} dx^i \tag{3.28}$$

where

$$\Delta x_{kl}^{0(mn)} = x_k^{0(m)} - x_l^{0(n)}. \tag{3.29}$$

with $k, l = 0, \cdots, 2$ and $m, n = 0, 1$. Then the CT time $x_1^{0(1)}$ at which the photon arrives in $p_1$ is given by $x_1^{0(1)} = x_0^{0(0)} + \Delta x_{10}^{0(10)}$. It is confirmed from (3.27) and (3.28) that $\Delta x_{20}^0 = 2 dl_\bullet$. The equations (3.27) and (3.28) clearly show that the speeds of light are anisotropic with respect to CT. The anisotropy is inherent as long as it is isotropic in $S$. In other words $\Delta x_{10}^{0(10)} \neq \Delta x_{21}^{0(01)}$ no matter what the definition of distance between $p_0$ and $p_1$ is.

In the GPS case, when making a first-order approximation, $dl_\bullet = dl$ and $g_{00}^{-1} g_{0i} dx^i = -\tilde{\beta} \tilde{r} d\tilde{\phi} = -d\tilde{\xi}$. Then

$$\Delta x_{10}^{0(10)} = dl + d\tilde{\xi} \tag{3.30}$$

$$\Delta x_{21}^{0(01)} = dl - d\tilde{\xi}. \tag{3.31}$$

Recall $dl = d\tilde{l}$. Equation (3.30) corresponds to (2.9). The one-way speed of light can be $c$ with respect to CT only if the Sagnac correction is absent from the time intervals. Its existence means the anisotropy of the speed of light.

## IV. ANALYSIS IN OBSERVATION SYSTEMS

Traditionally the Galilean transformation in cylindrical coordinates has been employed for a relativistic transformation between rotating and inertial frames without sufficient justification. Recently an exact transformation between rotating and inertial observation systems has been presented [9]. We here investigate speeds of light via the exact one.

### 1. Observation Systems and GPS

An observation system consists of world lines of observers. A coordinate vector in cylindrical coordinates including the time component can be represented as $\boldsymbol{p} = [ct, r, \phi, z]^T$ in an inertial observation system $S$ and as $\tilde{\boldsymbol{p}}' = [c\tilde{t}', \tilde{r}', \tilde{\phi}', \tilde{z}']^T$ in a rotating observation system $\tilde{S}'$. The inertial observation system $S$ is a collection of the world lines of inertial observers and is assumed to be isotropic. The rotating observation system $\tilde{S}'$ is composed of the world lines of rotating observers. The space-time coordinate transformation between $S$ and $\tilde{S}'$ can be written as [9]

$$\tilde{t}' = \gamma_\beta^{-1} t, \quad \tilde{r}' = h_\beta r, \quad \tilde{\phi}' = \varsigma_\beta \phi - \omega' \tilde{t}', \quad \tilde{z}' = z \tag{4.1}$$

where

$$\beta = c^{-1} r \omega \tag{4.2}$$

$$\varsigma_\beta = (1+\beta^2)^{-1/2} \tag{4.3}$$

$$h_\beta = \gamma_\beta^{-1} \varsigma_\beta^{-1} = (1-\beta^4) \tag{4.4}$$

and $\gamma_\beta$ is given as (2.5) with $\alpha = \beta$. The $\omega'$ is an angular frequency for an observer $\tilde{O}'$ located at radius $\tilde{r}'$ relative to $S'$, which is the primed observation system corresponding to the unprimed $S$. As for $\omega$, it is the unprimed angular frequency corresponding to $\omega'$. The coordinate transformation between $S$ and $S'$ is given by

$$t' = \gamma_\beta^{-1} t, \quad r' = h_\beta r, \quad \phi' = \varsigma_\beta \phi, \quad z' = z. \tag{4.5}$$

One can see from (4.1) and (4.5) that $\tilde{t}' = t'$ and $\tilde{r}' = r'$.

The world lines of observers are absolute in the sense that events on world lines such as time intervals represent real physical quantities, which cannot be changed by coordinate transformation although they can be seen differently in another coordinate system. The observation systems are composed of world lines and can be regarded as absolute in that sense.

In the case of the GPS, $S$ and $\tilde{S}'$ correspond to the ECI and ECEF frames, respectively. The coordinate system $\tilde{S}$ in Section II is similar to $\tilde{S}$, which is the unprimed observation system corresponding to $\tilde{S}'$. The observation system $\tilde{S}$ is rotating with angular frequency $\omega$ relative to $S$ which is identical with $S$ in Section II and the coordinate transformation between $S$ and $\tilde{S}$ can be written as (2.1). When an observer $\tilde{O}$, who is equivalent to the primed observer $\tilde{O}'$ seen in the unprimed system, is rotated by $\phi = \omega t$ relative to $S$, the observer $\tilde{O}'$ is rotated by $\phi' = \omega' t'$ relative to $S'$ where $\omega'$ and $\omega$ are related from (4.5) by

$$\omega' = \frac{d\phi'}{dt'} = \frac{d\phi'}{d\phi} \frac{d\phi}{dt} \frac{dt}{dt'} = h_\beta^{-1} \omega. \tag{4.6}$$

The normalized instantaneous speed of $\tilde{O}'$ relative to $S'$ is $\beta' = r'\omega'/c$. It is easy to see from (4.5) and (4.6) that $\beta' = \beta$. The angular frequency $\omega'$ is the angular speed of the Earth. The radius $r'$ can be written as $r' = R\cos\varphi$ where $\varphi$ is a latitude for the position of $\tilde{O}'$ and $R$ is the radius of the Earth.

In $\tilde{S}'$, the differential distance is calculated as $d\tilde{l}' = (dr'^2 + r'^2 d\tilde{\phi}'^2 + dz'^2)^{1/2}$, which means its spatial space is Euclidean. Within the first-order approximation, $\gamma_\beta = 1$, $\varsigma_\beta = 1$, and $h_\beta = 1$. Then (4.1) reduces to

$$\tilde{t}' = t, \quad \tilde{r}' = r, \quad \tilde{\phi}' = \phi - \omega'\tilde{t}', \quad \tilde{z}' = z. \tag{4.7}$$

The transformation (2.1) becomes equal to (4.7) if $\omega = \omega'$. Though (2.1) is nonrelativistic in itself, it, as a first-order approximation to (4.1), can be regarded as relativistic. Based on (4.1), the reason why (2.3), which is an approximate representation in $\tilde{S}'$ of $ds^2$, is used in the calculation for the GPS positioning can be explained as follows.

For simplicity suppose that the receiving position $\tilde{p}'_0$ as well as the transmitting position $\tilde{p}'_n$ in $\tilde{S}'$ is known. As mentioned above, $\tilde{S}'$ can be considered as absolute since it consists of world lines. The CT interval $\Delta\tilde{t}'_n$, which is the travel time in $\tilde{S}'$ of the signal transmitted from a satellite $A_n$, is a real quantity physically measured. Even though $\tilde{p}'_0$ and $\tilde{p}'_n$ are known, we cannot calculate $\Delta\tilde{t}'_n$ directly in $\tilde{S}'$ because the speed of light is unknown in $\tilde{S}'$. However we know that it is $c$ in $S$ so that $ds^2 = 0$. We can calculate $\Delta\tilde{t}'_n$ by using this fact. The $ds^2$ can be represented in $\tilde{S}'$ by using (4.1), and it is approximately expressed as (2.3) with appropriate notational change. Let the solution to the quadratic equation, $ds^2 = 0$, with respect to $d\tilde{t}'$ be $d\tilde{p}'$, which is a function of spatial coordinates $(\tilde{r}', \tilde{\phi}', \tilde{z}')$. The interval $\Delta\tilde{t}'_n$ is obtained by integrating $d\tilde{p}'$ along the path from $\tilde{p}'_n$ to $\tilde{p}'_0$. The $\Delta\tilde{t}'_n$ obtained is the actual physical time in $\tilde{S}'$.

Maybe, anybody would not think that the distance between a GPS satellite and a GPS receiver should be calculated as (3.24) in the ECEF frame and that AT, rather than CT, should be used as (3.3). However, someone may try to make the speed of light appear to be $c$ by introducing AT into $\tilde{S}'$. Unfortunately, as soon as AT is introduced, the Sagnac correction disappears, which is seen by substituting (2.9) or (3.27) with no approximation into (3.3). As a result of the cancellation, the apparent speed of light becomes $c$ with respect to AT if the spatial distance is measured as (3.24). Furthermore, there exists the problem of time gap that multiple AT times are defined at the same place [6, 8, 9] when the standard synchronization is set up along a closed path, for example a latitude, on the Earth. This matter is discussed in the next subsection.

## 2. Sagnac Effect

The ephemerides and the time information that the GPS satellites are broadcasting may contain some errors although they are so small that the effects on the positioning may be negligible. Though the computation of the speed of light in the GPS depends on such information, however, the analysis in Sagnac-type experiments is independent of man-made parameters, which can lead us to make more accurate analysis on constancy of the speed of light on a rotating plate.

The experiment of the Sagnac effect [10, 11] is performed on a rotating plate, which rotates around its center with an angular velocity of $\omega$. In the experiment, $S$ is a Laboratory frame and $\tilde{S}'$ a rotating frame. Recall $\beta' = \beta$. As seen in (4.4), $h_\beta$ is a function of $\beta$ or $\beta'$ and so is the ratio $\omega/\omega'\ (=h_\beta)$. Thus if $\omega\ (\omega')$ is independent of radius $r\ (r')$, $\omega'\ (\omega)$ depends on $r'\ (r)$. Though it cannot be ensured that $\omega$ is really constant regardless of $r$ in the Sagnac experiment, we assume that $\omega$ is a constant independent of $r$ and that the angular period of the rotating plate is $2\pi$ in $S$ so that the corresponding primed period in $S'$ becomes $2\pi\varsigma_\beta\ (<2\pi)$. The $\omega'$ in $\beta'\ (=r'\omega'/c)$, (4.6), and the third equation of (4.1) is the angular frequency defined with respect to the $2\pi$ revolution. Though the angular period for a turn in the primed observation system is less than $2\pi$, we use the $\omega'$, but keeping in mind that the angular period in $S'$ is $2\pi\varsigma_\beta$.

A light detector $\tilde{O}'$ is located at a radius $r_0$ from the center in $S$. At $t = t' = 0$, two light beams leave a light source, which is located at the same place as the detector, and begin to travel in the circular paths in opposite directions. The arrival times in $S$ of the co-rotating light beam $b_+$ and the counter-rotating one $b_-$ are given, respectively, by

$$t_{D\pm} = \frac{2\pi r_0}{c(1 \mp \beta)} \tag{4.8}$$

where $\beta = r_0\omega/c$. The relativistic difference between the arrival times is obtained as [9]

$$\Delta t'_d = \frac{4\pi r_0^2 \omega}{c^2(1-\beta^2)^{1/2}}. \tag{4.9}$$

.

## 2.1. Time differences and Speeds of Light

The time difference can be discovered by examining the motion of the light beams in the observation systems together with the coordinate transformation (4.1). In $S$, the rotation angles $\phi_{l\pm}$ of $b_\pm$ are given, respectively, by

$$\phi_{l\pm} = \pm\frac{ct}{r_0}. \tag{4.10}$$

The light beams traverse the circumference of radius $r_0$ in $\tilde{S}$, and radius $r'_0$ in $\tilde{S}'$ where

$$r_0' = h_\beta r_0. \tag{4.11}$$

Their rotation angles at an instant $t'$ in $S'$ are written, by using (4.10), (4.11), and the first and third equations in (4.5), as

$$\phi'_{l\pm} = \varsigma_\beta \phi_{l\pm} = \pm \frac{ct'}{r_0'}. \tag{4.12}$$

It is interesting that $\phi'_{l\pm}$ have the same forms as $\phi_{l\pm}$. When $b_\pm$ arrive at the detector, their rotation angles in $S$ are $\phi_{lD\pm} = \pm 2\pi/(1\mp\beta)$, which are expressed in $S'$ as

$$\phi'_{lD\pm} = \pm \frac{2\pi \varsigma_\beta}{1 \mp \beta}. \tag{4.13}$$

The angular frequencies of $b_\pm$ in $S'$ are given by

$$\omega'_{l\pm} = \frac{d\phi'_{l\pm}}{dt'} = \pm \frac{c}{r_0'}. \tag{4.14}$$

Using (4.13) and (4.14), the arrival times in the primed system are obtained as

$$t'_{D\pm} = \frac{2\pi \varsigma_\beta r_0'}{c(1\mp\beta)}. \tag{4.15}$$

Equation (4.15), with the use of (4.4), (4.8) and (4.11), can be rewritten as $t'_{D\pm} = t_{D\pm}/\gamma_\beta$, which is consistent with the first equation in (4.1) or (4.5). It is easy to see that the time difference $\Delta t'_d = t'_{D+} - t'_{D-}$ is given as (4.9).

Let us find the speeds of light in $S'$ and $\tilde{S}'$. When the light beams $b_\pm$ return to the detector, their rotation angles in $\tilde{S}'$ are given, using (4.13), (4.15), and $\beta' = r_0'\omega'/c\,(=\beta)$, by

$$\begin{aligned}
\tilde{\phi}'_{lD\pm} &= \phi'_{lD\pm} - \omega' t'_{D\pm} \\
&= \frac{2\pi \varsigma_\beta}{1\mp\beta}(\pm 1 - \beta) = \pm 2\pi \varsigma_\beta
\end{aligned} \tag{4.16}$$

The velocities of $b_\pm$ in $S'$ are calculated from (4.13) and (4.15) as

$$v'_\pm = \frac{\phi'_{lD\pm} r_0'}{t'_{D\pm}} = \pm c \tag{4.17}$$

and in $\tilde{S}'$

$$\tilde{v}'_\pm = \frac{\tilde{\phi}'_{lD\pm} r_0'}{t'_{D\pm}} = \pm c(1\mp\beta). \tag{4.18}$$

The speeds of $b_\pm$ are other than $c$ in $\tilde{S}'$ though they are $c$ in $S'$.

## 2.2. Standard Synchronization and No Fringe Shifts

To make the speed of light appear to be $c$ around a point $p_0$, it is necessary to adjust by (3.3) the coordinates of time at points infinitesimally near to $p_0$. When the time adjustment for light speed constancy is applied to closed paths, the problem of time gap is caused as the time to be adjusted, which will be called a time gap, depends on a closed path selected. The differential time gap is expressed as

$$dt'_{tg} = -c^{-1} g_{00}^{-1} g_{0i} dx^i. \tag{4.19}$$

Exploiting the time gap, the time difference in the experiment of the Sagnac effect can be discovered [11]. The proper time gap along a closed path is written as

$$t'_{\circ tg} = \int_{cp} |g_{00}|^{1/2} dt_{tg} \tag{4.20}$$

where *cp* stands for a closed path. For the paths $cp_\pm$ that the light beams $b_\pm$ traverse, the proper time gaps are given by

$$t'_{\circ tg \pm} = \pm 2\pi c^{-1} \beta \gamma_\beta r_0. \tag{4.21}$$

For derivation, refer to Appendix. It is straightforward to see that the difference $\Delta t'_{\circ tg} = t'_{\circ tg+} - t'_{\circ tg-}$ is calculated as (4.9).

Exploiting (4.1), we now analyze the Sagnac effect in the observation system $\tilde{S}'_{ss}$ with the standard synchronization for light speed constancy. In $\tilde{S}'_{ss}$, the travel distances $l'_{\bullet \pm}$ of $b_\pm$ are given by

$$l'_{\bullet +} = l'_{\bullet -} = \gamma_\beta^2 r'_0 \tilde{\phi}'_{lD} \tag{4.22}$$

and their arrival times $t'_{\bullet \pm}$ are obtained as

$$t'_{\bullet +} = t'_{\bullet -} = c^{-1} \gamma_\beta^2 r'_0 \tilde{\phi}'_{lD} \tag{4.23}$$

where $\tilde{\phi}'_{lD} = 2\pi \varsigma_\beta$. For details, see Appendix. Comparing (4.22) and (4.23), one can see that the speeds of $b_\pm$ are $c$, as expected. Since the light beams $b_\pm$ travel the same distance with the same apparent speed of $c$, their arrival times appear to be the same in $\tilde{S}'_{ss}$. It is easy to see that even if the metric tensor (3.7) obtained from the nonrelativistic transformation is applied, they travel the same distance. If the speeds of $b_\pm$ are really $c$, their arrival times are the same and the Sagnac effect cannot take place.

The reason the time difference of (4.9) can be obtained using the time gap of (4.19) is readily seen. From (3.3) and (4.19),

$$dt'_{CT} = dt'_{AT} + dt'_{tg} \tag{4.24}$$

where we use subscripts CT and AT to explicitly indicate related times. In the transformation between $S$ and $\tilde{S}'$, $\tilde{t}'(=t')$ includes time dilation factor $\gamma_\beta$ as in (4.1), and $g_{00} = -1$ so that $t'_{\circ tg} = t'_{tg}$. The CT interval is obtained by integrating (4.24) along a path given. As shown in (4.23), the AT intervals along $cp_\pm$ are the same. Thus the difference $\Delta t'_{tg}$ of time gap becomes equal to the CT difference $\Delta t'_{CT} = t'_{CT+} - t'_{CT-}$ where $t'_{CT\pm}$ are the CT intervals along $cp_\pm$. The Sagnac effect clearly indicates that the speeds of light are not equal. Meanwhile, the Sagnac correction in the GPS corresponds to $dt'_{tg}$ and the CT interval is obtained by integrating (4.24) from $\tilde{p}'_n$ to $\tilde{p}'_0$ where the AT interval is found by dividing by $c$ the integration of (3.24), which can be approximated as the Euclidean distance between $\tilde{p}'_n$ and $\tilde{p}'_0$. The speed of light appears to be $c$ with respect to AT. The Sagnac correction clearly indicates that the speed of light is anisotropic on the Earth.

From (4.23), one can explicitly see the standard synchronization and the time gap problem. Let us explain them in detail. The arrival times of (4.23) can be expressed as

$$t'_{\bullet+} = t'_{\bullet-} = \frac{1}{2}\left(\frac{\tilde{\phi}'_{ID} r'_0}{c(1-\beta)} + \frac{\tilde{\phi}'_{ID} r'_0}{c(1+\beta)}\right). \tag{4.25}$$

The light beams $b_\pm$ fully rotate the circumference in $\tilde{S}$ and thus their rotation angles are $\tilde{\phi}_{ID\pm} = \pm 2\pi$. If $\tilde{\phi}_{ID\pm} = \pm 2\pi$, $\tilde{\phi}'_{ID\pm} = \pm 2\pi \varsigma_\beta$, as shown in (4.16). In Fig. 3, the disk is rotating clockwise with angular frequency $\omega'$ in $S'$, the light source is located at $P_0$ in $\tilde{S}'$, and $P_\pm$ indicate the positions rotated by $\tilde{\phi}'_{ID\pm}$ from $P_0$. It should be noted that the figure has been drawn as if the rotation period of $\tilde{\phi}'$ were $2\pi$. Its period is $2\pi \varsigma_\beta$, as explained above. Then the spatial points $P_0$, $P_+$ and $P_-$ become the same one and the problem of time gap is clearly shown. Simply we can consider in terms of the synchronization that $P_0$ corresponds to $p_0$ and that $P_+$ or $P_-$ corresponds to $p_1$ in Fig. 2. Thus the AT time at $P_0$ is the same as CT, as can be seen from (3.14), while the AT time at $P_+$ or $P_-$ is given as (3.15). We cannot make a standard synchronization at $P_0$ since $P_0$, $P_+$ and $P_-$ are the same position.

Despite it, let us apply the standard synchronization to $P_+$ and $P_-$ as if they were different from $P_0$. The arc lengths between $P_0$ and $P_\pm$ are equal to $\tilde{\phi}'_{ID} r'_0$. In $\tilde{S}'$, according to (4.18), the speed of light is $c(1-\beta)$ in the rotation direction and $c(1+\beta)$ in the opposite direction. The $t'_{\bullet\pm}$ in (4.25) are half the roundtrip times for $cp_\pm$, which are the particular times for the standard synchronization. Even though such a synchronization may make the speeds of light appear to be $c$, the actual speeds of light are not changed by it. If the speeds of light are really $c$ in $\tilde{S}'$, there must be no fringe shifts in the Sagnac experiment, as shown above. Even if the standard synchronization is employed in $\tilde{S}'$ for light speed constancy, the actual speeds of $b_\pm$ are other than $c$ so that fringe shifts are observed. Moreover the introduction of the synchronization causes the time gap problem. It depends on a closed path selected. One can select an infinite number of different closed paths and then an infinite number of times can be defined at a spatial point.

## V. CONCLUSION

The GPS system has been known to be able to provide accurate information on GPS users' times and positions. Hence the GPS empirical measurements can show whether or not the one-way speed of light is actually isotropic. The navigation equations for GPS positioning may appear relativistic as they seem to exploit the invariance of space-time intervals together with the Langevin metric that has been usually used for the relativistic analysis of circular motion. However, the equations solve CT, not AT, under the Galilean transformation between the ECI and ECEF frames so that the speed of light with respect to CT inevitably becomes different from $c$ in the ECEF frame. To obtain accurate information from the GPS data, the Sagnac correction should be made. If the standard synchronization between spatially separated clocks is carried out, the term for the Sagnac correction is cancelled out. The GPS empirical measurements clearly show that the one-way speed is anisotropic in the ECEF frame. Besides, the problem of time gap is caused when the standard synchronization is applied to the ECEF frame.

The speed of light has been further investigated through the exact coordinate transformation between the inertial and rotating observation systems, $S$ and $\tilde{S}'$. The observation systems can be considered as absolute since they are composed of the world lines of observers. The Galilean transformation that the navigation equations employ can be justified as it is a first-order approximation to the exact one. This fact explains why the GPS positioning, in which the effects of the second and higher order terms are negligible, has been working so well under the nonrelativistic transformation. The Sagnac effect also clearly indicates that the speed of light is anisotropic in $\tilde{S}'$. It may be insisted that the speed of light can be $c$ by introducing the metric tensor. However, if the standard synchronization is applied to $\tilde{S}'$ so that the speed of light is really $c$, fringe shifts cannot occur because the two counter-rotating light beams travel the same distance. The occurrence of the Saganc effect as well as the need for the Sagnac correction is a clear indication of the anisotropy of the light speeds in rotating frames. Furthermore, the standard synchronization is impossible to set up along a closed path on account of the time gap problem, as explicitly shown.

We on the Earth live in a Euclidean space, not a curved space in which time and distance should be measured as (3.3) and (3.24) for the light speed constancy.

## APPENDIX

Consider that the metric tensor for the invariance of space-time intervals is employed in the observation system $\tilde{S}'$ and that $r$ is fixed and so is $r'$. Using (4.1), we have

$$dt = \gamma_\beta d\tilde{t}', \quad dr = 0, \quad d\phi = \varsigma_\beta^{-1}(d\tilde{\phi}' + \omega' d\tilde{t}'), \quad dz = 0. \tag{A.1}$$

Recall that $r' = \tilde{r}'$, $\tilde{t}' = t'$ and $\beta' = \beta$. Substituting (A.1) and $r = \gamma_\beta \varsigma_\beta r'$ into (2.2) yields

$$ds^2 = -c^2 dt'^2 + 2\gamma_\beta^2 \beta r' d\tilde{\phi}'(cdt') + \gamma_\beta^2 r'^2 d\tilde{\phi}'^2, \tag{A.2}$$

which leads to

$$g_{00} = -1, \quad g_{02} = g_{20} = \gamma_\beta^2 \beta r', \quad g_{22} = \gamma_\beta^2 r'^2 \tag{A.3}$$

where $(x^0, x^1, x^2) = (c\tilde{t}', \tilde{r}', \tilde{\phi}')$. Note that AT equals PT and $dl_\bullet = dl_\circ$ since $g_{00} = -1$. In the Sagnac experiment, $r = r_0$ and $r' = r'_0$.

From (4.19) and (A.2)

$$dt'_{tg} = c^{-1} \gamma_\beta^2 \beta r'_0 d\tilde{\phi}'. \tag{A.4}$$

Using $t' = \gamma_\beta^{-1} t$ and $\omega' = \gamma_\beta \varsigma_\beta \omega$, we have $\omega' t' = \varsigma_\beta \omega t$. The third equation in (4.1) can be expressed as $\tilde{\phi}' = \varsigma_\beta \phi - \omega' t' = \varsigma_\beta \tilde{\phi}$ and then

$$d\tilde{\phi}' = \varsigma_\beta d\tilde{\phi}. \tag{A.5}$$

Putting (A.5) and $r'_0 = \gamma_\beta^{-1} \varsigma_\beta^{-1} r_0$ in (A.4), it follows that

$$dt'_{tg} = c^{-1} \gamma_\beta \beta r_0 d\tilde{\phi}. \tag{A.6}$$

Since $|g_{00}| = 1$, $dt'_{\circ tg} = dt'_{tg}$. Integration of (A.6) form 0 to $\pm 2\pi$ with respect to $\tilde{\phi}$ yields (4.21).

The spatial distances $l'_{\bullet\pm}$ that $b_\pm$ traverse in the observation system $\tilde{S}'_{ss}$ with the standard synchronization for light speed constancy can be discovered using (3.24). Inserting (A.3) in (3.4) with $r' = r'_0$, it follows that

$$\begin{aligned} dl'_\bullet &= \gamma_\beta r'_0 \, (1 + \gamma_\beta^2 \beta^2)^{1/2} |d\tilde{\phi}'| \\ &= \gamma_\beta^2 r'_0 \, |d\tilde{\phi}'| \end{aligned}. \tag{A.7}$$

As seen in (4.16), $\tilde{\phi}'$ is equal to $\tilde{\phi}'_{lD\pm}$ at $t' = t'_{D\pm}$. Integrating (A.7) along the paths that $b_\pm$ traverse, we have

$$l'_{\bullet\pm} = \int_0^{|\tilde{\phi}'_{lD\pm}|} \gamma_\beta^2 r'_0 d\tilde{\phi}' = \gamma_\beta^2 r'_0 |\tilde{\phi}'_{lD\pm}|. \tag{A.8}$$

The arrival times $t'_{\bullet\pm}$ of $b_\pm$ in $\tilde{S}'_{ss}$ can be expressed, from (3.3) and (4.19), as

$$t'_{\bullet\pm} = \int_0^{t'_{D\pm}} (dt' - dt'_{tg}) = t'_{D\pm} - t'_{tg\pm}. \tag{A.9}$$

Using (A.4) and recalling that the light beams $b_\pm$ rotate in $\tilde{S}'$ by $\tilde{\phi}'_{lD\pm}$ when they return to the detector, $t'_{tg\pm}$ are given by

$$t'_{tg\pm} = \int_0^{\tilde{\phi}'_{lD\pm}} c^{-1} \gamma_\beta^2 \beta r'_0 d\tilde{\phi}' = c^{-1} \gamma_\beta^2 \beta r'_0 \tilde{\phi}'_{lD\pm}. \tag{A.10}$$

One can confirm by inserting (4.11) and (4.16) in (A.10) that the resultant is equal to (4.21). Noting $\tilde{\phi}'_{lD+} = 2\pi\varsigma_\beta$, $t'_{D\pm}$ of (4.15) is rewritten as

$$t'_{D\pm} = \frac{r'_0 \tilde{\phi}'_{lD+}}{c(1 \mp \beta)}. \tag{A.11}$$

Recalling $\gamma_\beta^2 = 1/(1-\beta^2)$, substitution of (A.10) and (A.11) into (A.9) yields (4.23).

**FIGURE CAPTIONS**

Fig. 1. CT and AT over world lines.

Fig. 2. Roundtrip of an object and standard synchronization at $p_1$.

Fig. 3. Standard-synchronization-associated closed paths between $P_0$ and $P_+$ and between $P_0$ and $P_-$.

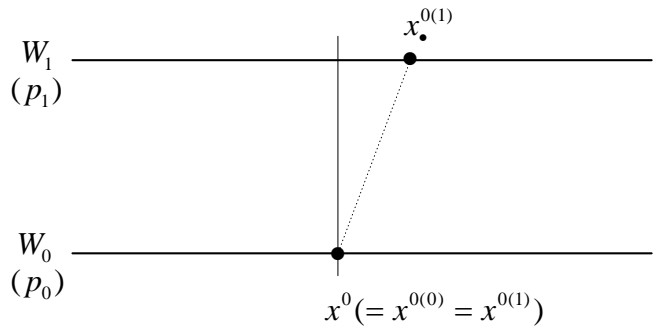

Fig. 1. CT and AT over world lines.

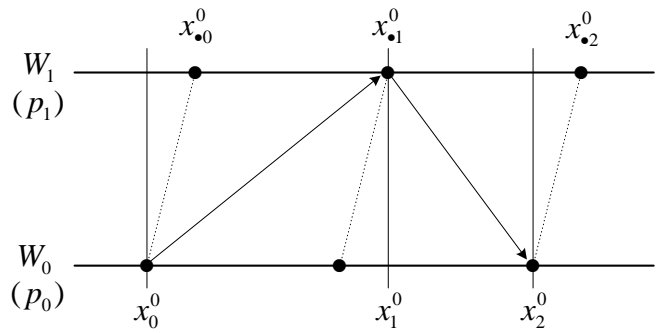

Fig. 2. Roundtrip of an object and standard synchronization at $p_1$.

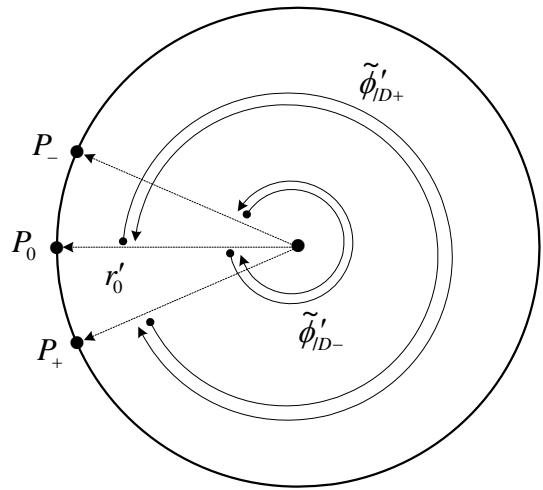

Fig. 3. Standard-synchronization-associated closed paths between $P_0$ and $P_+$ and between $P_0$ and $P_-$.